\documentclass{aastex}          
\usepackage{spr-astr-addons}    

\begin{document}
%
\title{A new photometric and spectroscopic study of the eclipsing binaries CC~Her and CM~Lac: Physical parameters and evolutionary status}

\shorttitle{Photometric and Spectroscopic of Analysis CC~Her and CM~Lac}

\shortauthors{A. Liakos \& P. Niarchos}

\author{A. Liakos\altaffilmark{}}
\email{alliakos@phys.uoa.gr} 
\and
\author{P. Niarchos\altaffilmark{}}
\affil{Department of Astrophysics, Astronomy and Mechanics, National and Kapodistrian University of Athens, GR 157 84, Zografos, Athens, Hellas}


\begin{abstract}
New complete light and radial velocities curves were obtained for the eclipsing binaries CC~Her and CM~Lac. The data are analysed with modern techniques in order to derive the physical parameters of the systems and study their present evolutionary status. We found that CC~Her is a classical Algol type binary, while CM~Lac is a detached system with two Main Sequence stars in asynchronous orbit.
\end{abstract}


\section{Introduction}
\label{INTRO}

The main purpose of the present study is the derivation of the physical parameters and estimation of the evolutionary status of the components of the eclipsing double--line spectroscopic binaries CC~Her and CM~Lac. The present study is based on new photometric and spectroscopic data. The systems presented herein were chosen for the following reasons: (1) Their existing light curves (hereafter LCs) were either incomplete, or not multi-filtered, (2) they are bright enough to obtain radial velocities measurements (hereafter RVs), (3) they are candidates for triplicity \citep{HO06}, and (4) they are also candidates for including an oscillating component \citep{SOE06}. New RVs and LCs of these systems were obtained with the aim: (a) to derive the absolute elements of their components, (b) to search photometrically and spectroscopically for the presence of a tertiary companion and (c) to look for possible pulsational behaviour of the components.

\textbf{CC~Her} is an Algol type binary with a period of 1.73405$^{\rm d}$ and a maximum magnitude in $V$-filter of 9.95~mag \citep{WA06}. \citet{KR71} noticed for the first time that orbital period changes were occurring in the system, while \citet{SOF06} performed an orbital period analysis and they found that a third body with mass of 2.69~$M_\odot$ might orbit the eclipsing pair. Many spectral classifications have been made for the system ranging between A0-A2 spectral types \citep[cf.][]{MA06,HA84,FA02}, but it seems that the spectral type has been settled as A2 (SIMBAD).

\textbf{CM~Lac} has 1.60469$^{\rm d}$ period and a maximum $V_{\rm mag}$ of 8.18 \citep{WA06}. \citet{RO56} classified the primary component as A7 type star and \citet{BA68} mentioned that the system consists of two Main Sequence stars of A2 and A8 spectral types. \citet{PO68} obtained the RVs of both components and estimated their masses as 1.87$\cdot\sin^3i~M_\odot$ and 1.47$\cdot\sin^3i~M_\odot$ for primary and secondary, respectively. \citet{OL84} calculated the rotational velocities of the components and proposed an asynchronous orbit. \citet{LI73} calculated the absolute elements of the system, while \citet{AA76} suggested a third light contribution to the LC solution. Although there are many observations for this system, the temperatures of the components are not determined accurately. However, SIMBAD, based on the \textsl{MK Classifications of Spectroscopic Binaries} of \citet{AB09}, refers the system as A3 type.

\section{Observations and data reduction}
\label{OBS}

\begin{table*}
\centering
\caption{The observations log.
\label{tab1}}
\begin{tabular}{lcc ccc}
\tableline
System   &     Obs. Dates    &           \multicolumn{4}{c}{Comparison stars}                                   \\
\tableline
                                    \multicolumn{6}{c}{Photometry}                                              \\
\tableline
         &                   &          $C$       &$V_{\rm mag}^{\rm a}$&      $K$      & $V_{\rm mag}^{\rm a}$ \\
\tableline
CC Her   & 13/04/09-16/06/09 &     TYC 946-103-1     &    10.05        &   TYC 946-1286-1    &   11.20          \\
CM Lac   & 17/08/09-08/09/09 &     TYC 3210-925-1    &    8.58         &   TYC 3210-1381-1   &   11.20          \\
\tableline
                                             \multicolumn{6}{c}{Spectroscopy}                                   \\
\tableline
         &                   &          star         &  spectral type  &          star       & spectral type    \\
\tableline
CC Her   &     12-14/05/09   &      HIP 53910        &       A1        &    HIP 70400        &       A5         \\
         &                   &      HIP 77233        &       A3        &    HIP 89935        &       A7         \\
CM Lac   & 3, 5, 6/10/10     &      HIP 6193         &       A3        &    HIP 23871        &       A5         \\
         &                   &      HIP 4283         &       A4        &    HIP 116928       &       A7         \\
\tableline
\multicolumn{6}{l}{$^{\rm a}$Taken from the Tycho-2 Catalogue \citep{HO00}}
\end{tabular}
\end{table*}

Spectroscopic observations were obtained with the 1.3~m Ritchey-Cretien telescope at Skinakas Observatory, Crete Is., Hellas, on May 2009 for CC~Her and on October 2010 for CM~Lac. A 2000$\times$800 ISA SITe CCD camera attached to a focal reducer with a 2400~lines/mm grating and slit of 80~$\mu$m was used. This arrangement gave a nominal dispersion of 0.55~\AA/pixel and wavelength coverage between 4534-5622~\AA~for CC~Her and 4775-5858~\AA~for CM~Lac. The spectral region was selected so as to include H$_{\beta}$ and sufficient metallic lines (e.g. MgI triplet). Data reduction was performed using the \emph{\textbf{Ra}dial \textbf{Ve}locity \textbf{re}ductions} v.2.1d software \citep{NE09}. The frames were bias subtracted, a flat field correction was applied, and the sky background was removed. Before and after each on-target observation, an arc calibration exposure (NeHeAr) was recorded.

The photometric observations were carried out at the Gerostathopoulion Observatory of the University of Athens during 23 nights (13 for CC~Her and 10 for CM~Lac) on April-September 2009, using the 0.4~m Cassegrain telescope equipped with the ST-10XME CCD camera and the $BVRI$ Bessell photometric filters. Aperture photometry was applied in the data and differential magnitudes were obtained using the software \emph{MuniWin} v.1.1.26 \citep{HR98}. For CC~Her, the time resolution of the data (i.e. the time difference between two successive images in the same filter) was $\sim3$~min, and the mean photometric error (measured in mmag) for each filter's data was: 2.8 (in $B$), 3.1 (in $V$), 3.6 (in $R$) and 3.3 (in $I$), while for CM~Lac were: $\sim50$~sec, 2.3 (in $B$), 2.2 (in $V$), 2.6 (in $R$) and 2.6 (in $I$), respectively. Further details on the comparison ($C$) and check stars ($K$) used for each programme are given in Table~\ref{tab1}.

\section{Data analysis}
\label{DA}

\subsection{Spectroscopic analysis}
\begin{figure}[h]
\centering
\includegraphics[width=8cm]{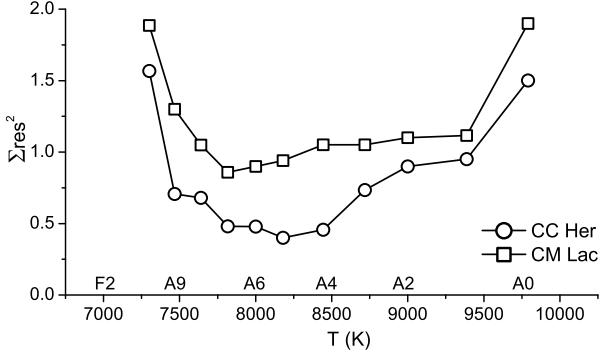}
\caption{Spectral type search plot for each system.}
\label{fig1}
\end{figure}

\begin{figure*}
\centering
\begin{tabular}{cc}
CC Her                                &             CM Lac                    \\
\includegraphics[width=8cm]{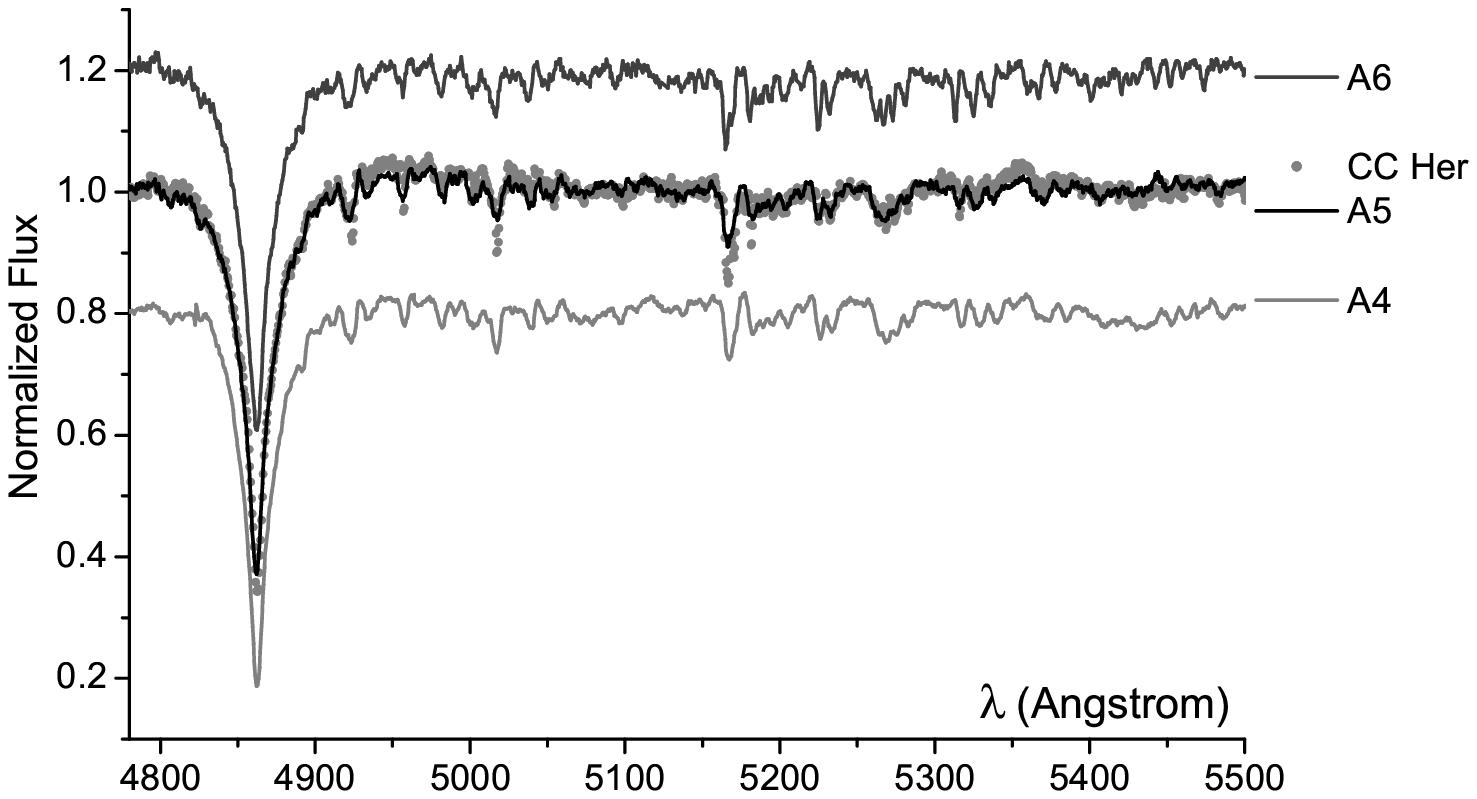}&\includegraphics[width=8cm]{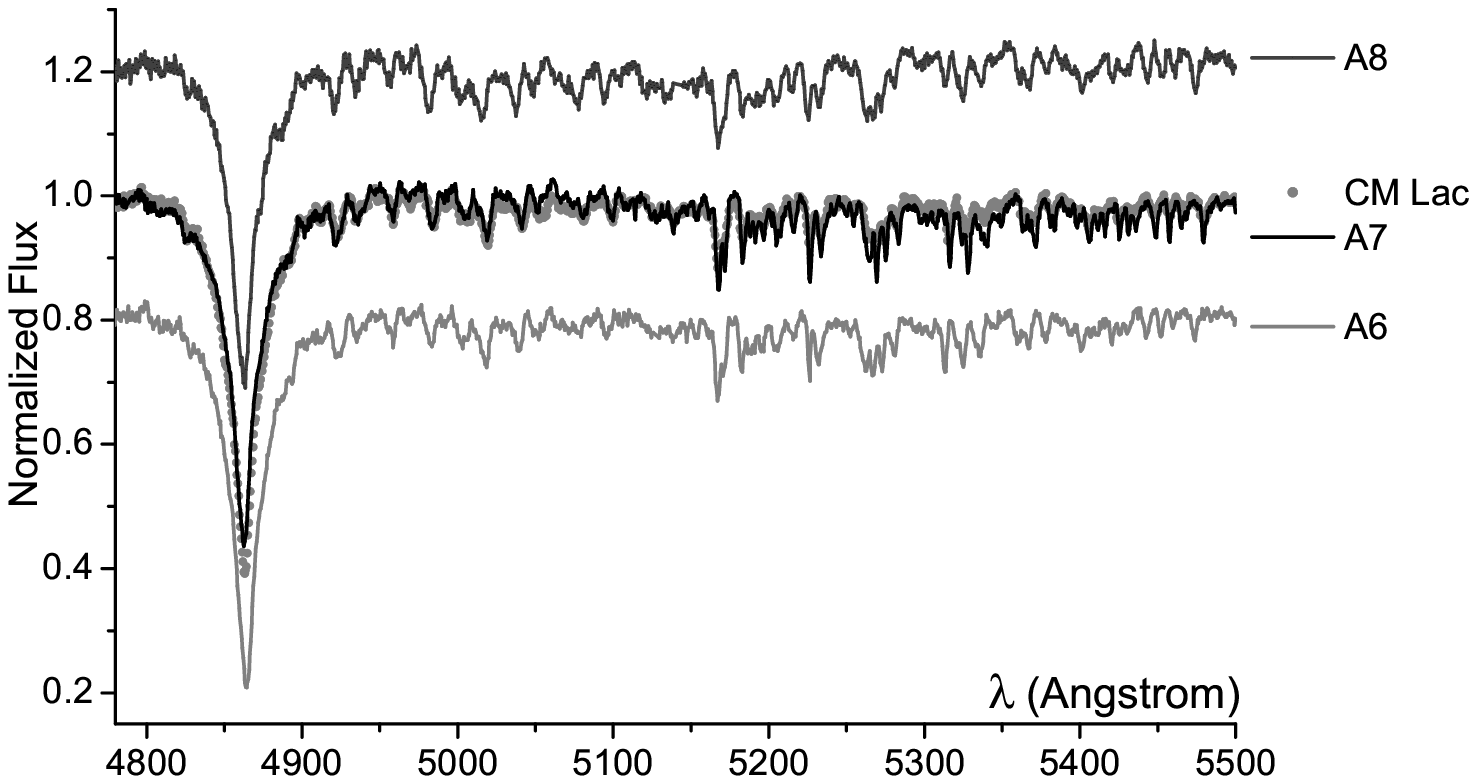} \\
\includegraphics[width=8cm]{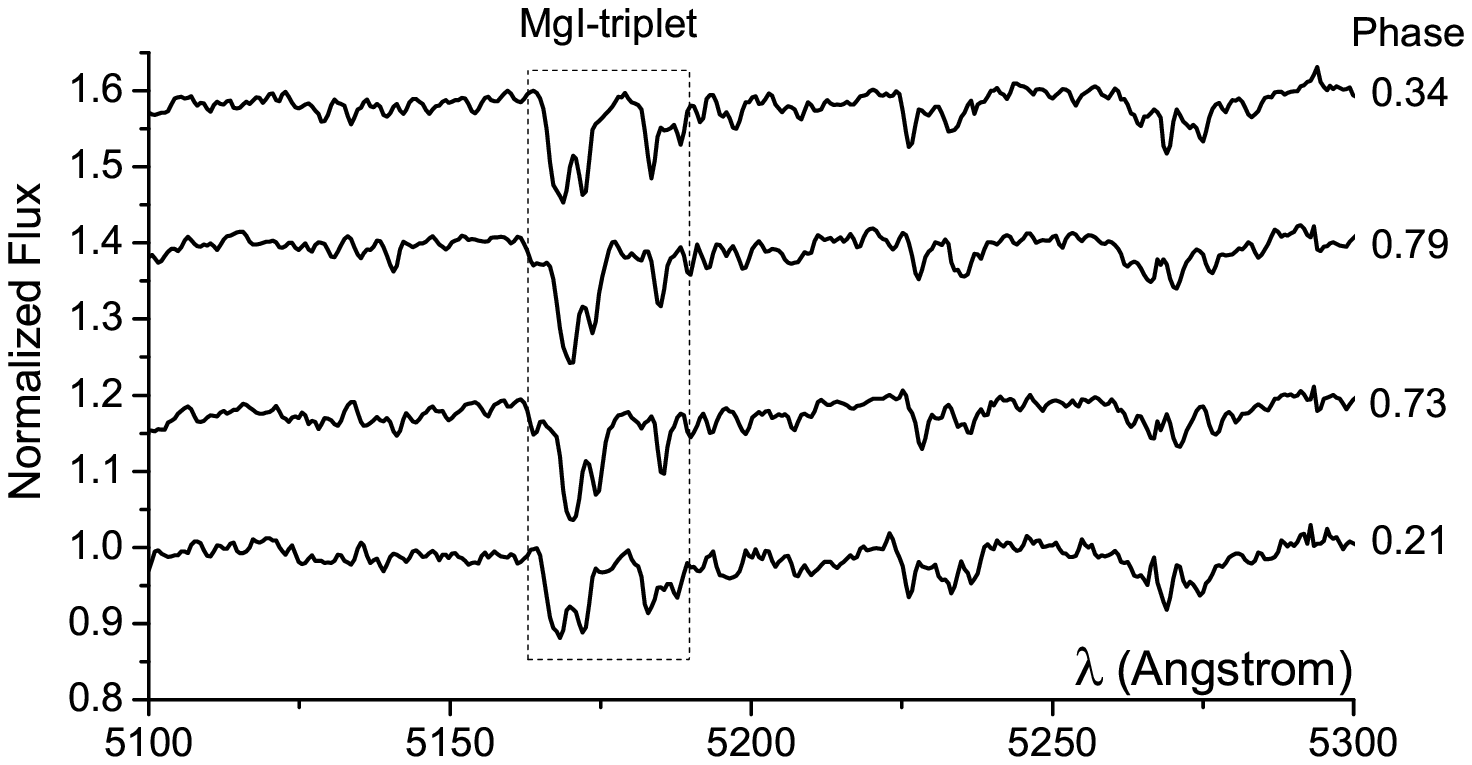}&\includegraphics[width=8cm]{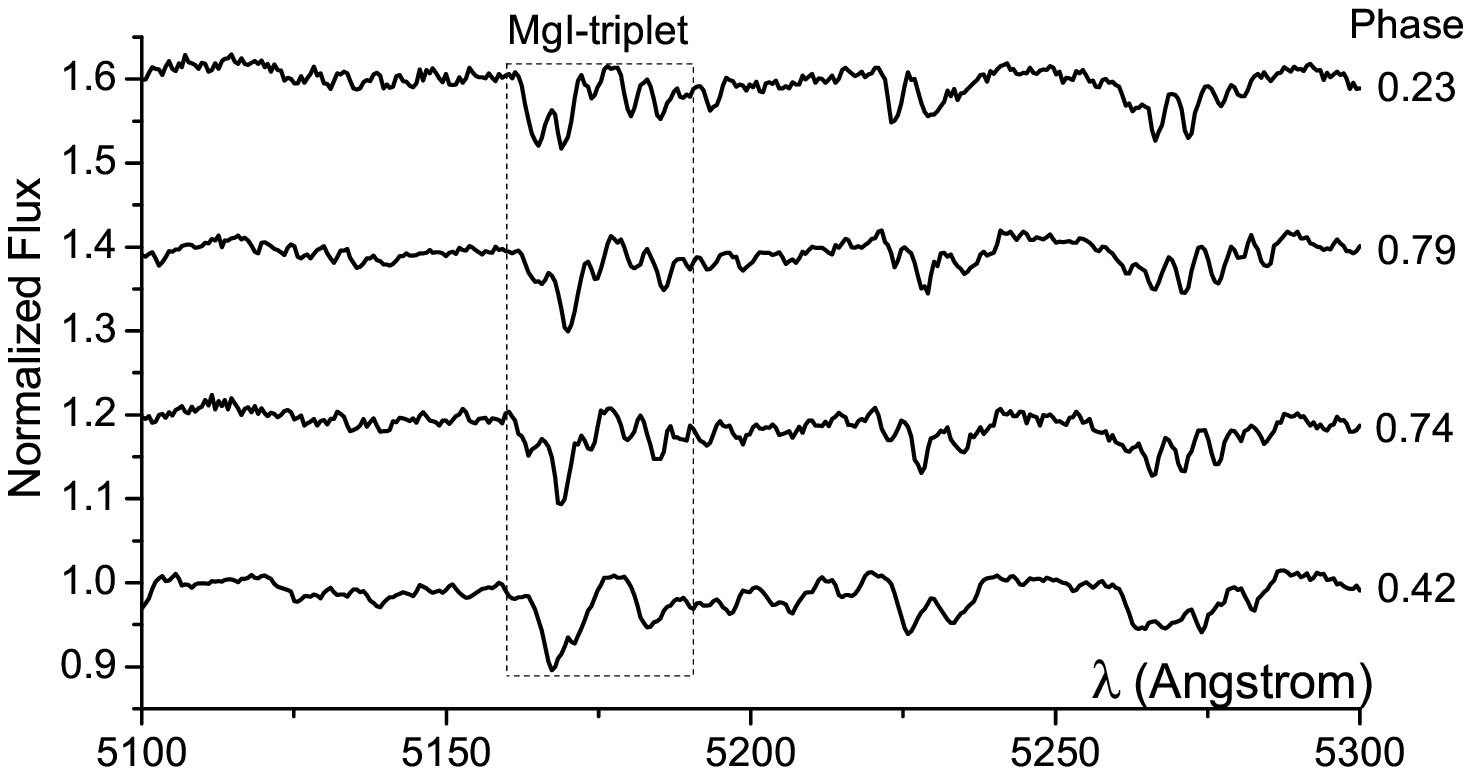}
\end{tabular}
\caption{Comparison spectra of the systems with standard stars (upper) and the motion of MgI triplet in various phases (lower).}
\label{fig2}
\end{figure*}

A total of 19 spectroscopic standard stars, suggested by \emph{GEMINI Observatory} (http://www.gemini.edu/), ranging from A0 to G8 spectral types were observed with the same instrumental set-up. Exposure times for the variables were 1800~sec for CC~Her and 900~sec for CM~Lac. All spectra were calibrated and normalized to enable direct comparisons. The spectral region between 4800~\AA~and 5350~\AA, where H$_{\beta}$ and numerous metallic lines are strong, was used for the spectral classification. The rest part of the spectra was ignored, because not enough metallic lines with significant signal-to-noise ratios existed. Due to the lack of spectroscopic observations during the eclipses, it was not possible to estimate the spectral type of the components of the systems, but only their combined spectral types. However, the combined spectra of the systems were compared with those of standard stars. The variables' spectra, taken near the maximum separation of the components (around phases 0.25 and 0.75), were subtracted from those of each standard deriving sums of squared residuals in each case (see Fig.~\ref{fig1}). Such least squares sums allowed the best match between the spectra of variable and standard to be found. For CC~Her the best comparison was found with HIP~70400 which is an A5 type star, while the combined spectrum of CM~Lac was fitted better with the spectrum of an A7 type star, namely HIP~116928. The estimated error for this method is one subclass. Comparison spectra are plotted in Fig.~\ref{fig2} for each system.

For the RVs calculations, the \emph{Broadening Functions} (BFs) method \citep{RU92,RU02} on the spectra was used. This method is based on the comparison between non-broadened lines, but perhaps with a small shift (i.e. the lines of a standard star), and broadened and shifted ones (i.e. the lines of a binary system). It produces both the broadening and shifting, that practically correspond to the RVs of the components. Briefly, the main difference between the BFs and the widely used Cross-Correlation-Function (CCF), as presented by \citep{RU99}, is the production of the baselines. The CCF, in the cases of binary spectra, indeed takes into account the shape of the spectrum, but does not exclude the natural broadening from the sharp-line template (i.e. sharp metallic lines). The latter defines poorly the baseline and the peak-pulling for the components. On the other hand, the BFs are based on linear equations, produce well-defined baselines, while the peak-pulling effects are absent.

We cropped all spectra in order to avoid the broad H$_{\beta}$ line, and we included all the sharp metallic lines between 4865-5355~\AA. Each RV value and its error was derived statistically (mean value and error) from the respective velocities resulted from this method by using different comparison standard stars with similar spectral types (see Table~\ref{tab1}). The heliocentric RVs are given in Table~\ref{tab2}, while a sample of spectra showing the metallic lines' motion during the orbital phase for each case are illustrated in Fig.~\ref{fig2}.

\begin{table*}
\centering
\caption{Heliocentric radial velocities measurements with their errors in parentheses for CC~Her and CM~Lac.
\label{tab2}}

\begin{tabular}{cccc cccc}
\tableline
    HJD	    &	Phase	&	  $RV_1$	&	   $RV_2$	&	   HJD	    &	Phase	&	  $RV_1$	&	  $RV_2$    \\
(2400000.0+)&           &    (km/sec)   &     (km/sec)  &  (2400000.0+) &           &    (km/sec)   &    (km/sec)   \\
\tableline
                                                  \multicolumn{8}{c}{CC~Her}                                        \\
\tableline
54964.5518	&	0.208	&	   -69 (31) &	156 (46)	&	54966.4147	&	0.282	&	-66 (26)	&	170 (68)    \\
54965.4666	&	0.735	&	   41 (27)	&	-185 (21)	&	54966.4472	&	0.301	&	-62 (18)	&	170 (47)    \\
54965.5007	&	0.755	&	   37 (18)	&	-194 (37)	&	54966.4792	&	0.319	&	-65 (31)	&	163 (74)    \\
54965.5291	&	0.771	&	   34 (9)	&	-198 (29)	&	54966.5293	&	0.348	&	-60 (11)	&	165 (61)    \\
54965.5619	&	0.790	&	   36 (8)	&	-197 (37)	&		        &		    &		        &		        \\
\tableline
                                                 \multicolumn{8}{c}{CM~Lac}                                         \\
\tableline
55473.2793	&	0.374	&	-64 (24)	&	153 (21)	&	55476.2638	&	0.233	&	-89 (20)	&	189 (20)	\\
55473.2912	&	0.381	&	-78 (17)	&	142 (16)	&	55476.2812	&	0.244	&	-91 (22)	&	184 (19)	\\
55473.3038	&	0.389	&	-63 (23)	&	140 (18)	&	55476.2892	&	0.249	&	-92 (19)	&	183 (21)	\\
55473.3167	&	0.397	&	-57 (27)	&	127 (30)	&	55476.2973	&	0.254	&	-87 (21)	&	190 (22)	\\
55473.3294	&	0.405	&	-36 (34)	&	106 (36)	&	55476.3054	&	0.259	&	-96 (20)	&	177 (21)	\\
55473.3420	&	0.413	&	-34 (19)	&	    --	    &	55476.3137	&	0.265	&	-96 (21)	&	173 (21)	\\
55473.3546	&	0.421	&	-26 (23)	&	    --	    &	55476.3436	&	0.283	&	-106 (21)	&	160 (25)	\\
55475.4708	&	0.739	&	147 (26)	&	-123 (25)	&	55476.3522	&	0.289	&	-100 (22)	&	170 (22)	\\
55475.4847	&	0.748	&	148 (22)	&	-126 (26)	&	55476.3609	&	0.294	&	-88 (21)	&	181 (22)	\\
55475.5039	&	0.760	&	128 (28)	&	-141 (20)	&	55476.3753	&	0.303	&	-91 (23)	&	173 (22)	\\
55475.5175	&	0.768	&	130 (27)	&	-137 (21)	&	55476.4529	&	0.351	&	-75 (23)	&	145 (25)	\\
55475.5311	&	0.777	&	142 (23)	&	-130 (21)	&	55476.4652	&	0.359	&	-81 (24)	&	142 (27)	\\
55475.5447	&	0.785	&	150 (21)	&	-127 (18)	&	55476.4789	&	0.368	&	-77 (24)	&	131 (26)	\\
55476.2630	&	0.233	&	-91 (21)	&	187 (19)	&	55476.4934	&	0.377	&	-86 (25)	&	131 (21)	\\
\tableline
\end{tabular}
\end{table*}

\subsection{Light and radial velocities curves analysis}

\begin{figure*}[t]
\centering
\begin{tabular}{cc}
CC Her                                 &            CM Lac                    \\
\includegraphics[width=8cm]{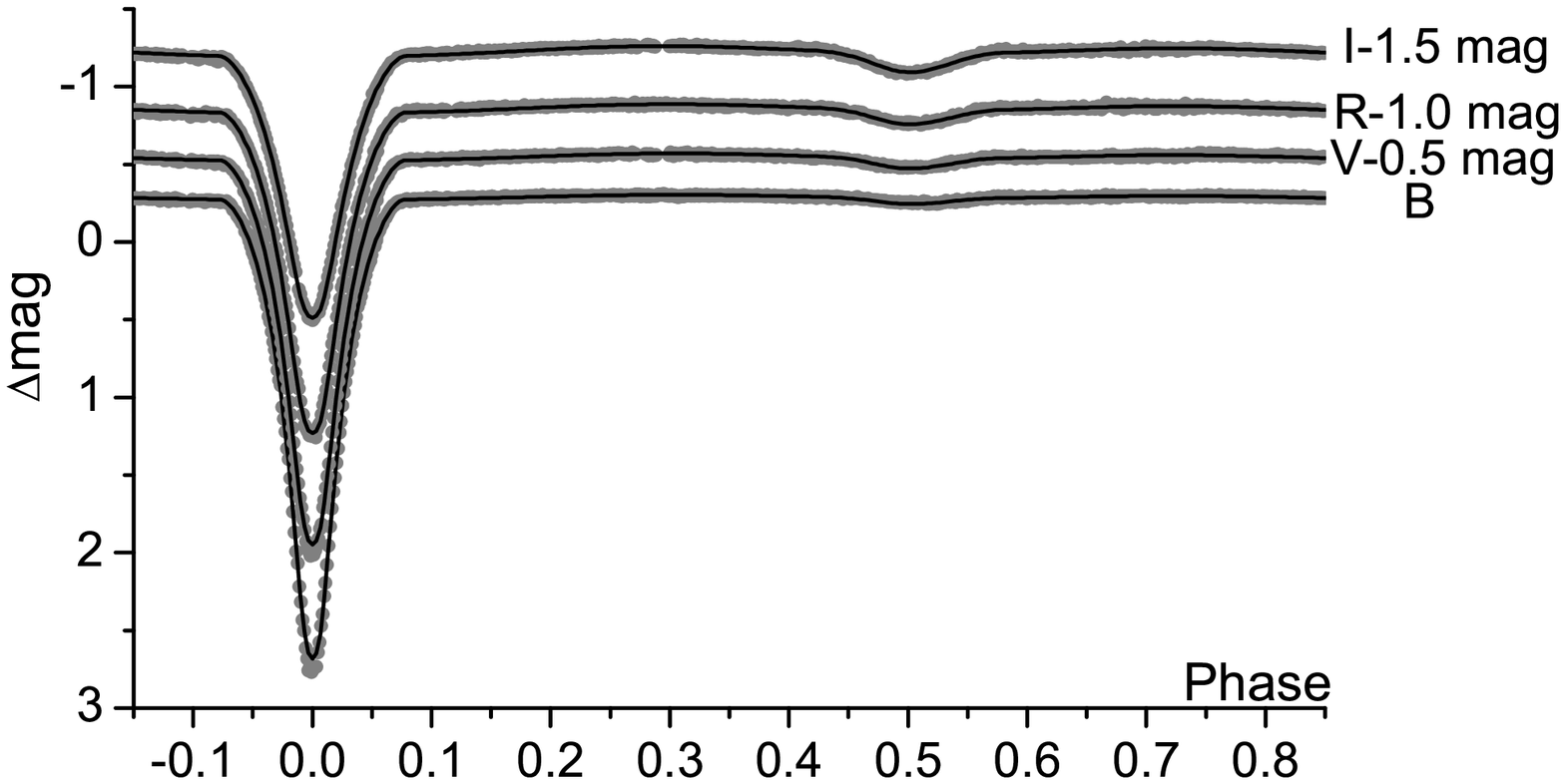} &\includegraphics[width=8cm]{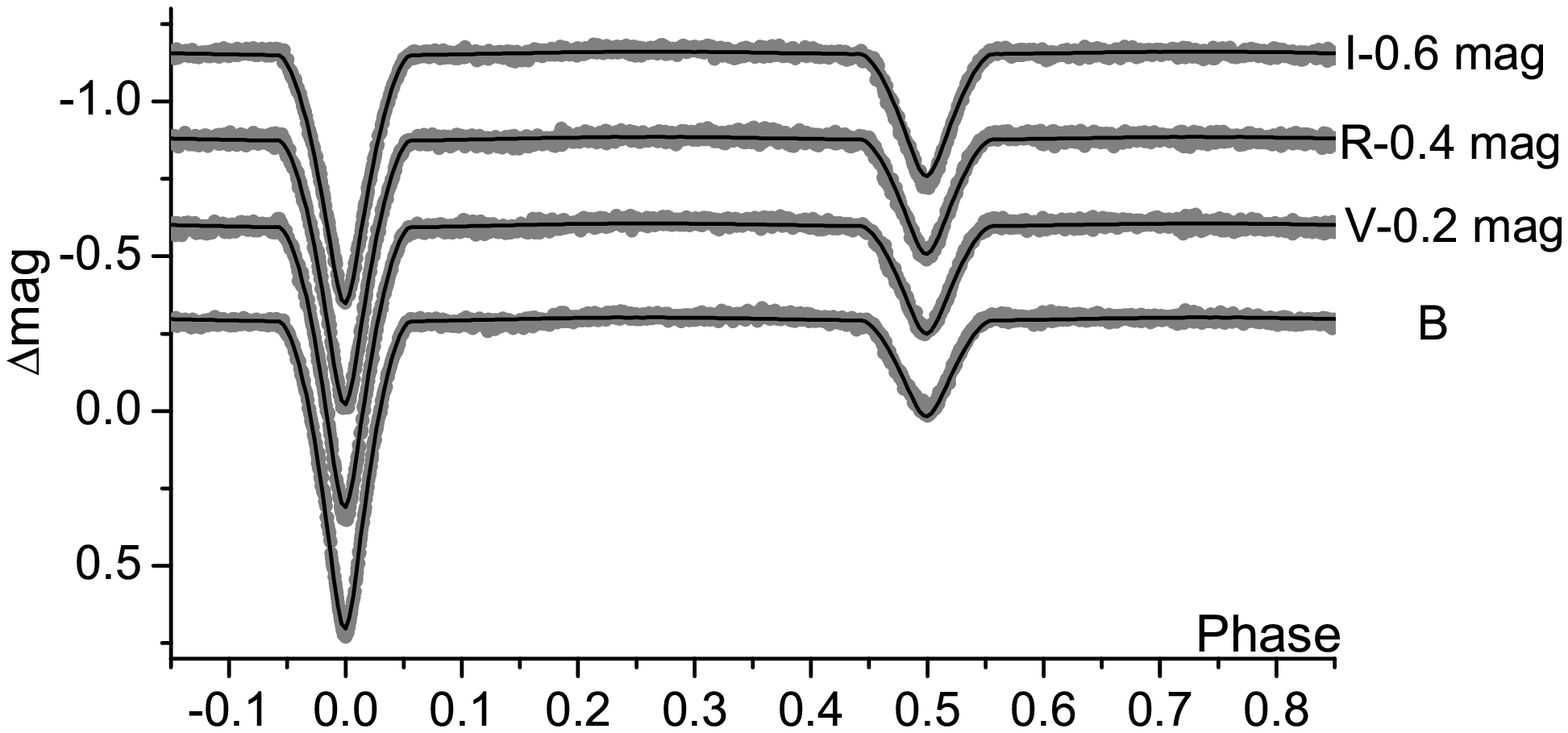}\\
\includegraphics[width=8cm]{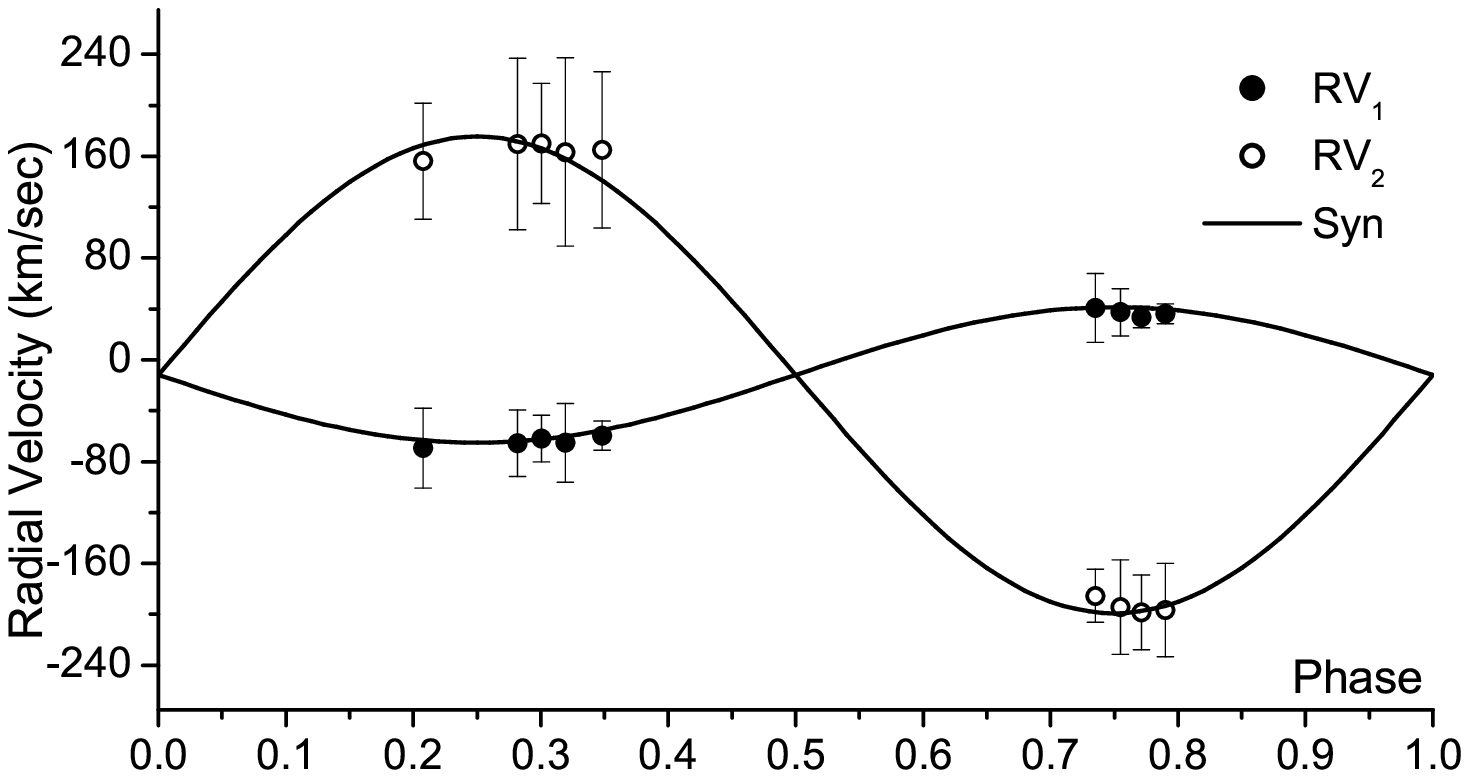} &\includegraphics[width=8cm]{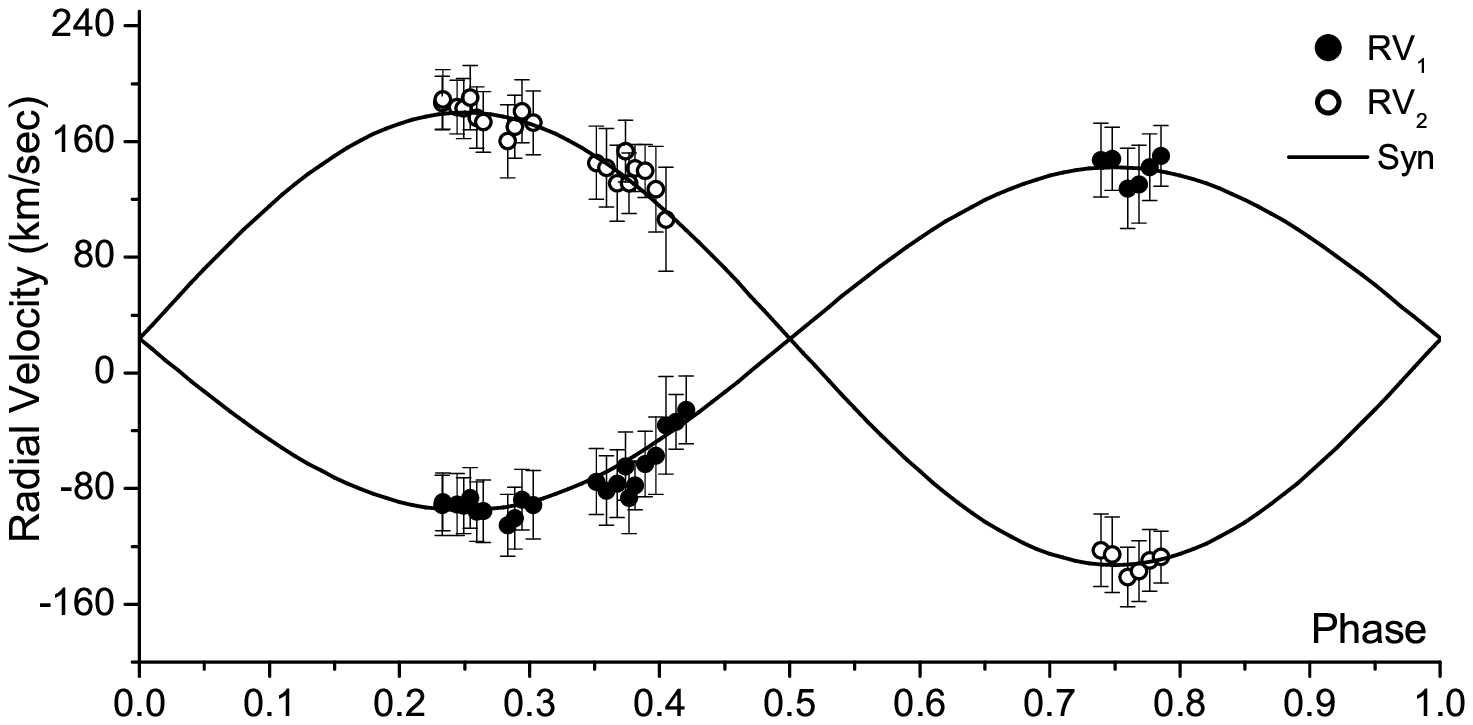}\\
\includegraphics[width=6.7cm]{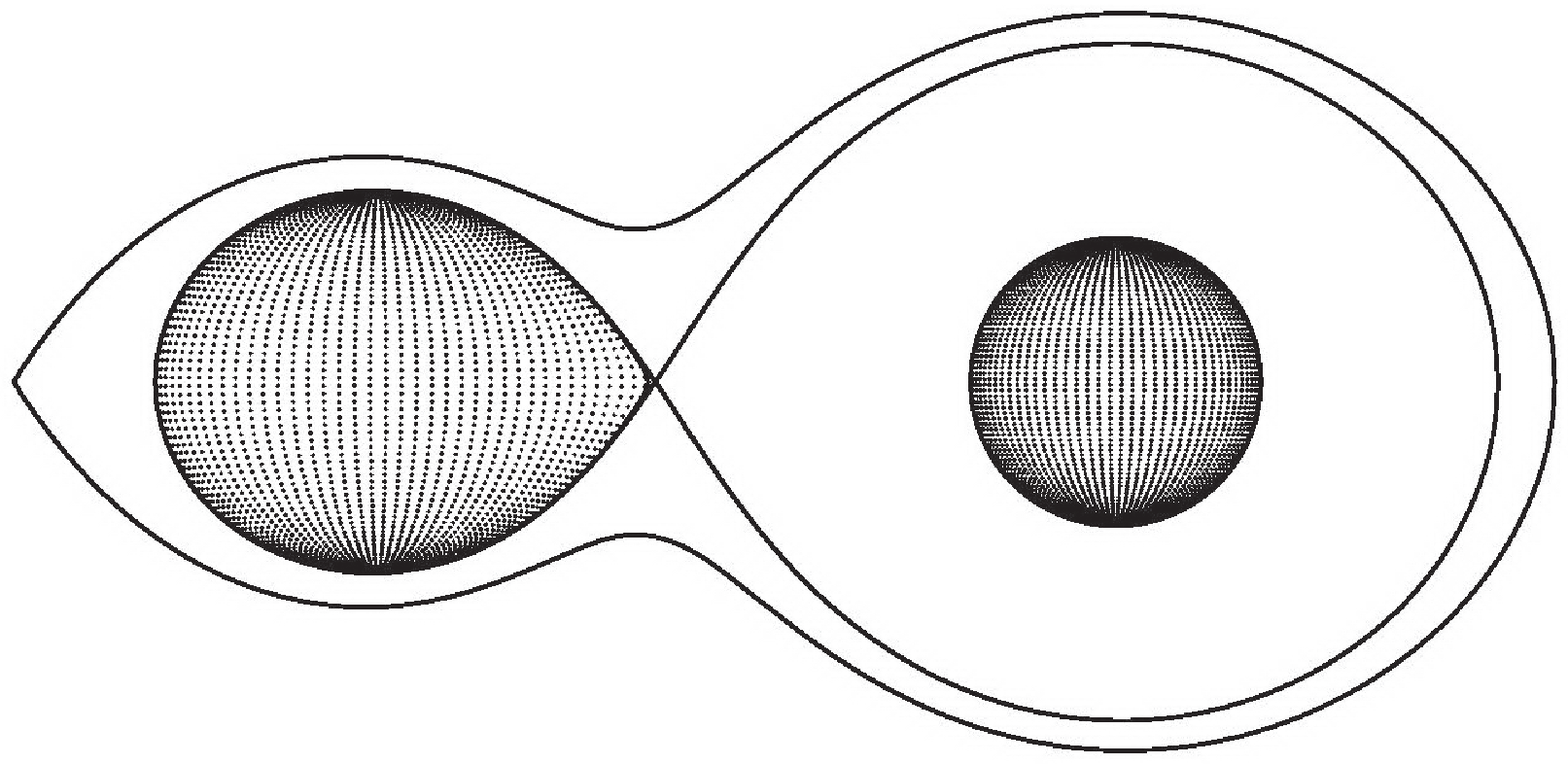} &\includegraphics[width=7cm]{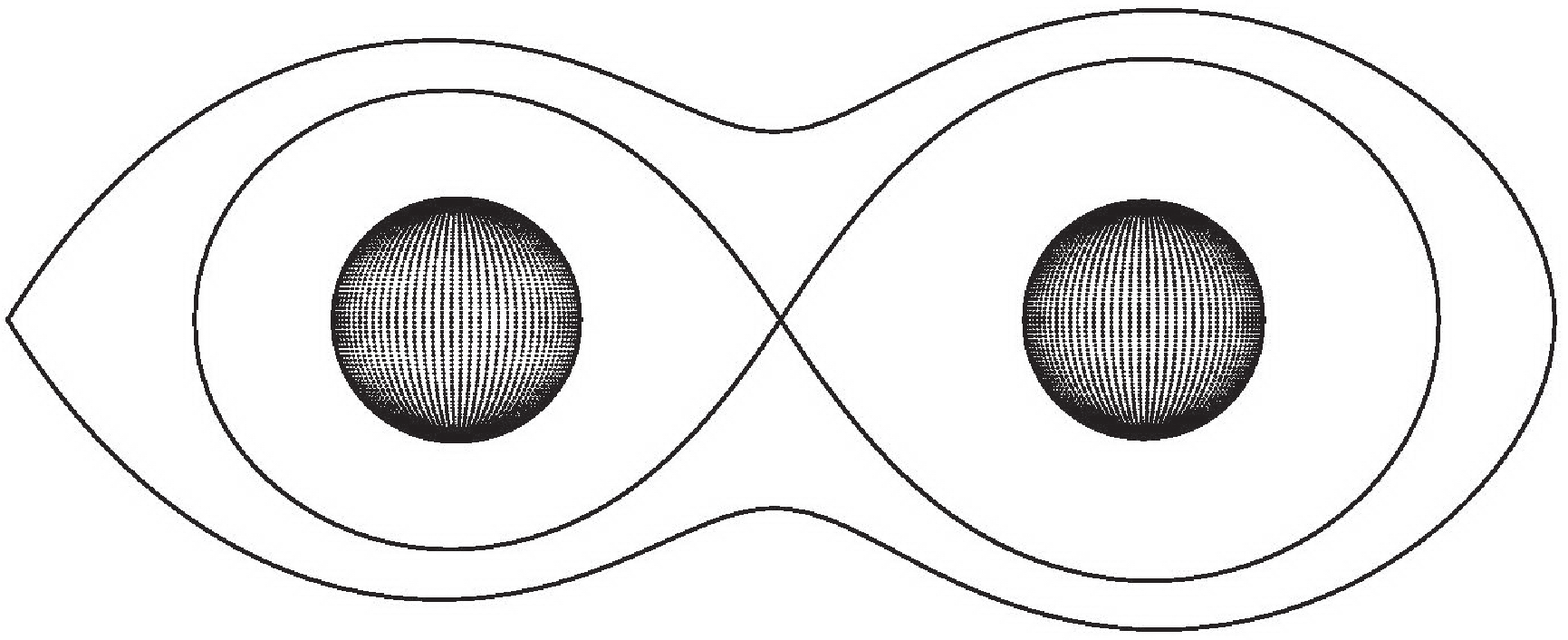}\\
\end{tabular}
\caption{Synthetic (solid lines) and observed (points) light (upper panels) and radial velocities curves (middle panels) and the 3D plots (lower panels) of the systems at the phase 0.75.}
\label{fig3}
\end{figure*}

The LCs and RVs were analysed simultaneously using the \emph{PHOEBE} v.0.29d software \citep{PZ05} which uses the method of the 2003 version of the Wilson-Devinney (WD) code \citep{WD71,WI79,WI90}. Modes 2 (detached system) and 5 (conventional semi-detached system) were chosen for fitting. The temperatures of the primaries are not determined rigorously for both systems. Literature contains controversial values, so we preferred to present models in a range of primaries' temperatures. However, the present spectral classifications can be used as the lower limit for the primary's temperature, since they were derived from spectra in which the luminosity contribution of the secondary component is maximum (see Section 3.1). Therefore, despite the light domination of the primary, the spectra are affected from the secondary resulting in more `cool' spectral types. On the other hand, as an upper temperature edge can be used other spectral classifications (see Section 1) of the systems. The temperature of the primary of each case was assigned value according to its spectral class \citep{CO00} and was kept fixed during the analysis, while the temperature of the secondary $T_2$ was adjusted. The albedos, $A_1$ and $A_2$, and gravity darkening coefficients, $g_1$ and $g_2$, were set to the generally adopted values for the given spectral types of the components \citep{VZ24,LU67,RU69}. The (linear) limb darkening coefficients, $x_1$ and $x_2$, were taken from the tables of \citet{VH93}. The dimensionless potentials $\Omega_{1}$ and $\Omega_{2}$, the synchronization parameters $F_1$ and $F_2$, the mass ratio $q$, the semi-major axis $a$, the systemic radial velocity $\gamma$, the fractional luminosity of the primary component $L_{1}$, the inclination $i$ of the system's orbit were set in the programme as adjustable parameters. In addition, due to literature information for possible triplicity \citep{HO06}, the third light option $l_{3}$ was also trialled. The semi-amplitudes of the RVs (maximum radial velocity of each component), $K_1$ and $K_2$, were calculated by fitting a sinusoidal function on the points of each curve. Best-fit models and the observed LCs and RVs, as well as the 3D plots of the systems are presented in Fig.~\ref{fig3} with corresponding parameters given in Table~\ref{tab3}.

\begin{table*}
\centering
\caption{Light and radial velocities curves modelling parameters. The errors are given in parentheses and correspond to the last digit.
\label{tab3}}
\begin{tabular}{l|ccc|ccc}
\tableline
Parameter                 &             \multicolumn{3}{c}{CC Her}      &               \multicolumn{3}{c}{CM Lac}        \\
\tableline
	                      &	$T_1$=9000 K&	$T_1$=8600 K&	$T_1$=8200 K&	$T_1$=8700 K&	$T_1$=8200 K&	$T_1$=7800 K  \\
\tableline
$i~(^\circ)$	          &	86.1 (1)	&	86.1 (1)	&	86.0 (1)	&	86.9 (1)	&	86.8 (1)	&	86.7 (1)	  \\
$q~(m_{2}/m_{1}$)	      &	0.29 (1)	&	0.29 (1)	&	0.29 (1)	&	0.76 (2)	&	0.76 (2)	&	0.76 (2)	  \\
$K_1$ (km/sec)	          &	54 (1)	    &	54 (1)	    &	54 (1)	    &	119 (2)	    &	119 (2)	    &	119 (2)	      \\
$K_2$ (km/sec)	          &	185 (2)	    &	185 (2)	    &	185 (2)	    &	156 (2)	    &	156 (2)	    &	156 (2)	      \\
$\gamma$ (km/sec)	      &	-12 (2)	    &	-12 (2)	    &	-12 (2)	    &	24 (1)	    &	24 (1)	    &	24 (1)	      \\
$a$ ($R_\odot$)	          &	8.2 (1)	    &	8.2 (1)	    &	8.2 (1)	    &	8.9 (1)	    & 	8.9 (1)	    &	8.9 (1)	      \\
$T_2$ (K)	              &	4586 (5)	&	4473 (5)	&	4360 (5)	&	7034 (7)	&	6705 (6)	&	6448 (6)	  \\
$\Omega_{1}$	          &	5.41 (1)	&	5.43 (1)	&	5.42 (1)	&	6.55 (1)	&	6.57 (1)	&	6.62 (1)	  \\
$\Omega_{2}$	          &	   2.45	    &	2.45	    &	  2.45	    &	5.40 (1)	&	5.39 (1)	&	5.36 (1)	  \\
$A_1$	                  &	  1$^a$	    &	1$^a$	    &	    1$^a$	&	    1$^a$	&	    1$^a$	&	1$^a$	      \\
$A_2$	                  &	0.75 (1)	&	0.82 (1)	&	0.88 (1)	&	 0.5$^a$	&	  0.5$^a$	&	0.5$^a$	      \\
$g_1$	                  &	  1$^a$	    &	1$^a$	    &	    1$^a$	&	    1$^a$	&	    1$^a$	&	1$^a$	      \\
$g_2$	                  &	0.32$^a$	&	0.32$^a$	&	0.32$^a$	&	0.32$^a$	&	0.32$^a$	&	0.32$^a$	  \\
$F_1$	                  &	  1$^a$	    &	1$^a$	    &	    1$^a$	&	    1.1 (2)	&	  1.1 (2)	&	0.8 (2)	      \\
$F_2$	                  &	  1$^a$	    &	1$^a$	    &	    1$^a$	&	0.88 (4)	&	 1.0 (3)	&	1.4 (1)	      \\
$x_{1,{\rm B}}$	          &	  0.538	    &	    0.559	&	    0.584	&	    0.547	&	    0.581	&	0.596	      \\
$x_{1,{\rm V}}$	          &	  0.458	    &	    0.480	&	    0.504	&	    0.471	&	    0.502	&	0.524	      \\
$x_{1,{\rm R}}$	          &	  0.380	    &	    0.402	&	    0.424	&	    0.396	&	    0.424	&	0.447	      \\
$x_{1,{\rm I}}$	          &	  0.299	    &	    0.319	&	    0.341	&	    0.316	&	    0.341	&	0.365	      \\
$x_{2,{\rm B}}$	          &	  0.939	    &	    0.956	&	    0.976	&	    0.599	&	    0.628	&	0.655	      \\
$x_{2,{\rm V}}$	          &	  0.790	    &	    0.805	&	    0.823	&	    0.494	&	    0.513	&	0.531	      \\
$x_{2,{\rm R}}$	          &	  0.679	    &	    0.696	&	    0.713	&	    0.415	&	    0.434	&	0.451	      \\
$x_{2,{\rm I}}$	          &	  0.565	    &	    0.578	&	    0.589	&	    0.338	&	    0.358	&	0.375	      \\
$(L_1/L_{\rm T})_{\rm B}$ &	0.933 (1)	&	0.932 (1)	&	0.932 (1)	&	0.703 (2)	&	0.698 (2)	&	0.686 (2)	  \\
$(L_1/L_{\rm T})_{\rm V}$ &	0.890 (1)	&	0.889 (1)	&	0.889 (1)	&	0.667 (2)	&	0.662 (2)	&	0.649 (2)	  \\
$(L_1/L_{\rm T})_{\rm R}$ &	0.839 (1)	&	0.838 (1)	&	0.838 (1)	&	0.645 (2)	&	0.640 (1)	&	0.627 (1)	  \\
$(L_1/L_{\rm T})_{\rm I}$ & 0.772 (1)	&	0.771 (1)	&	0.771 (1)	&	0.623 (1)	&	0.617 (1)	&	0.604 (1)	  \\
$(L_2/L_{\rm T})_{\rm B}$ &	0.056 (1)	&	0.056 (1)	&	0.056 (1)	&	0.297 (1)	&	0.302 (1)	&	0.314 (1)	  \\
$(L_2/L_{\rm T})_{\rm V}$ &	0.103 (1)	&	0.104 (1)	&	0.104 (1)	&	0.333 (1)	&	0.338 (1)	&	0.351 (1)	  \\
$(L_2/L_{\rm T})_{\rm R}$ &	0.144 (1)	&	0.145 (1)	&	0.146 (1)	&	0.355 (1)	&	0.360 (1)	&	0.373 (1)	  \\
$(L_2/L_{\rm T})_{\rm I}$ &	0.196 (1)	&	0.198 (1)	&	0.199 (1)	&	0.377 (1)	&	0.383 (1)	&	0.396 (1)	  \\
\tableline
                          &                             \multicolumn{6}{c}{Third light}                                   \\
\tableline
$(L_3/L_{\rm T})_{\rm B}$ &	0.011 (1)	&	0.011 (1)	&	0.011 (1)	&	      --	&	   --	    &	   --	      \\
$(L_3/L_{\rm T})_{\rm V}$ &	0.007 (1)	&	0.008 (1)	&	0.007 (1)	&	      --	&	   --	    &	   --	      \\
$(L_3/L_{\rm T})_{\rm R}$ &	0.017 (1)	&	0.017 (1)	&	0.016 (1)	&	      --	&	   --	    &	   --	      \\
$(L_3/L_{\rm T})_{\rm I}$ &	0.032 (2)	&	0.031 (2)	&	0.031 (2)	&	      --	&	   --	    &	   --	      \\
\tableline
                          &                             \multicolumn{6}{c}{Cool spot}                                     \\
\tableline
latitude $(^\circ)$	      &	  89 (2)	&	   89 (2)	&	   89 (2)	&	   --	    &	       --	&	   --	      \\
longitude $(^\circ)$	  &	 339 (2)	&	  339 (2)	&	  339 (2)	&	   --	    &	      --	&	   --	      \\
Radius $(^\circ)$		  &	   15 (2)	&	   15 (2)	&	   15 (2)	&	   --	    &	       --	&	   --	      \\
$T_{\rm {spot}}/T_{\rm {surf.}}$&0.80 (3)&	 0.80 (3)	&	 0.80 (3)	&	   --	    &	     --	    &	   --	      \\
\tableline
$\Sigma res^2$	          &	0.860	    &  	   0.863	&	    0.882	&	    0.043	&	    0.044	&	0.045	      \\
\tableline
\multicolumn{7}{l}{$^a$assumed, $L_{\rm T}=L_1+L_2+L_3$}
\end{tabular}
\end{table*}

For CC~Her, a range between A2-A5 type temperature values for the primary was chosen. Finally, a semidetached configuration was found to describe the LC better inside the selected range of temperatures. A small third light contribution was revealed in the LCs, but not also in the RVs. This discrepancy is further discussed in the last section. In addition, slight LC asymmetries were detected on the brightness maxima (Max.II-Max.I=0.01~mag in $I$-filter). For this reason, a cool spot on the secondary's surface was placed (O'Connell effect) and its parameters (latitude, longitude, radius and temperature factor) were also adjusted. The albedo $A_2$ of the secondary (possible reflection effect due to large temperature difference between the components) was left as free parameter for a better approach. Alternatively, since the secondary component fills its Roche Lobe it is a potential mass loser. This mass flow could impact the primary's surface (mass gainer) and cause brightness increase, therefore we also assumed a hot photospheric spot on it, instead of the cool spot scenario. The comparison between these two hypotheses, based on the sum of the squared residuals, showed that is more probable the existence of the cool spot on the cooler component of the system than the hot spot on the primary. In particular, for the model with $T_1=9000$~K, we found $\Sigma res^2$=0.860 for the cool spot scenario, while for the hot spot assumption this value turned out 0.935. The comparison of these values shows that the cool-spot hypothesis provides $\sim10\%$ better fitting than the hot-spot scenario. Almost the same differences resulted from the other models that based on different temperature of the primary.

The temperature of CM~Lac's primary was assigned values in the range A3-A7. Best fit was achieved in the detached mode. A third light was neither found in the LCs nor in the RVs, so we plausibly conclude that the triplicity scenario should be rejected.

\section{Absolute parameters of the components}
\label{ABS}
\begin{figure}[t]
\begin{tabular}{c}
\includegraphics[width=7.8cm]{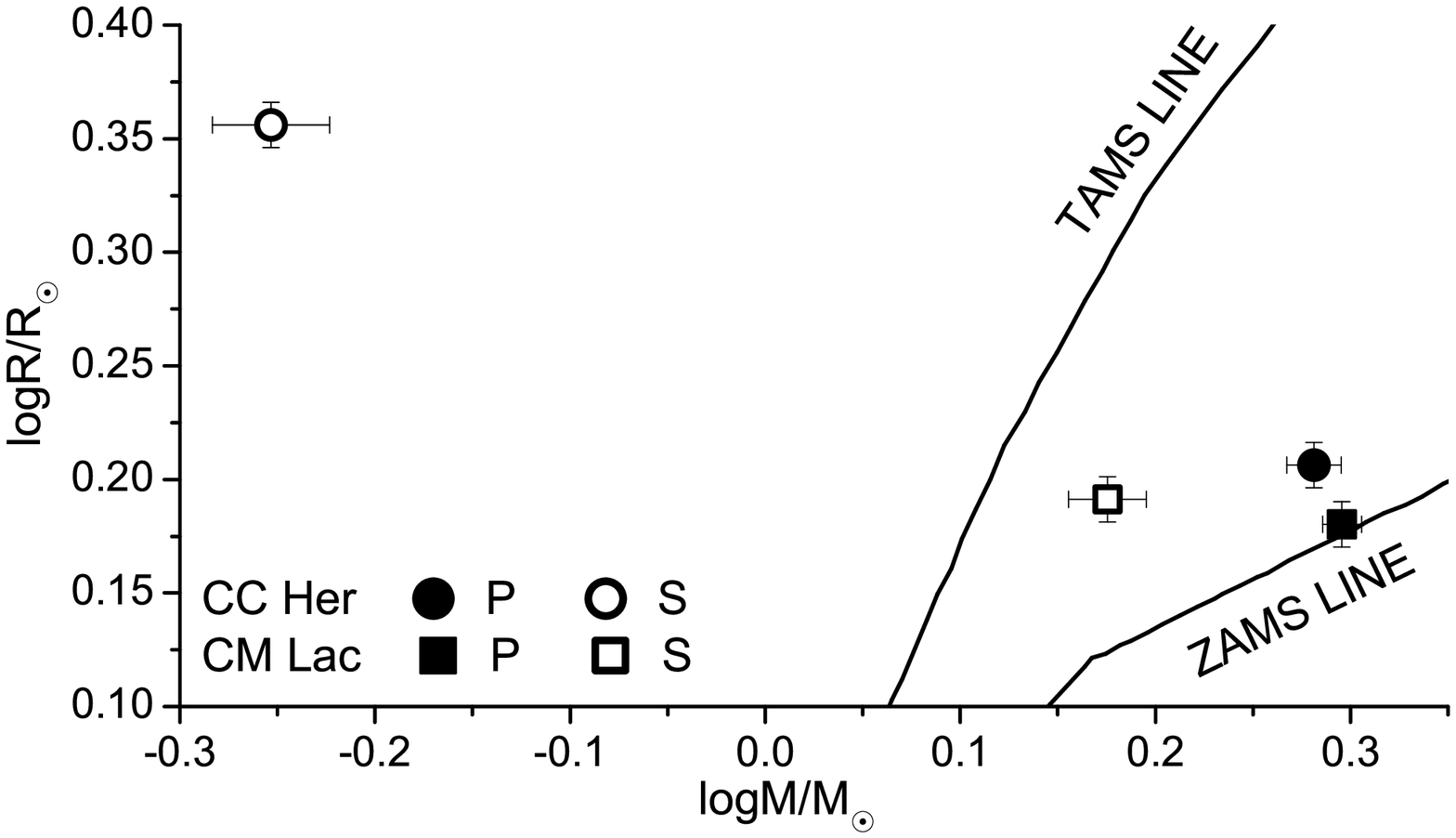}\\
\includegraphics[width=7.8cm]{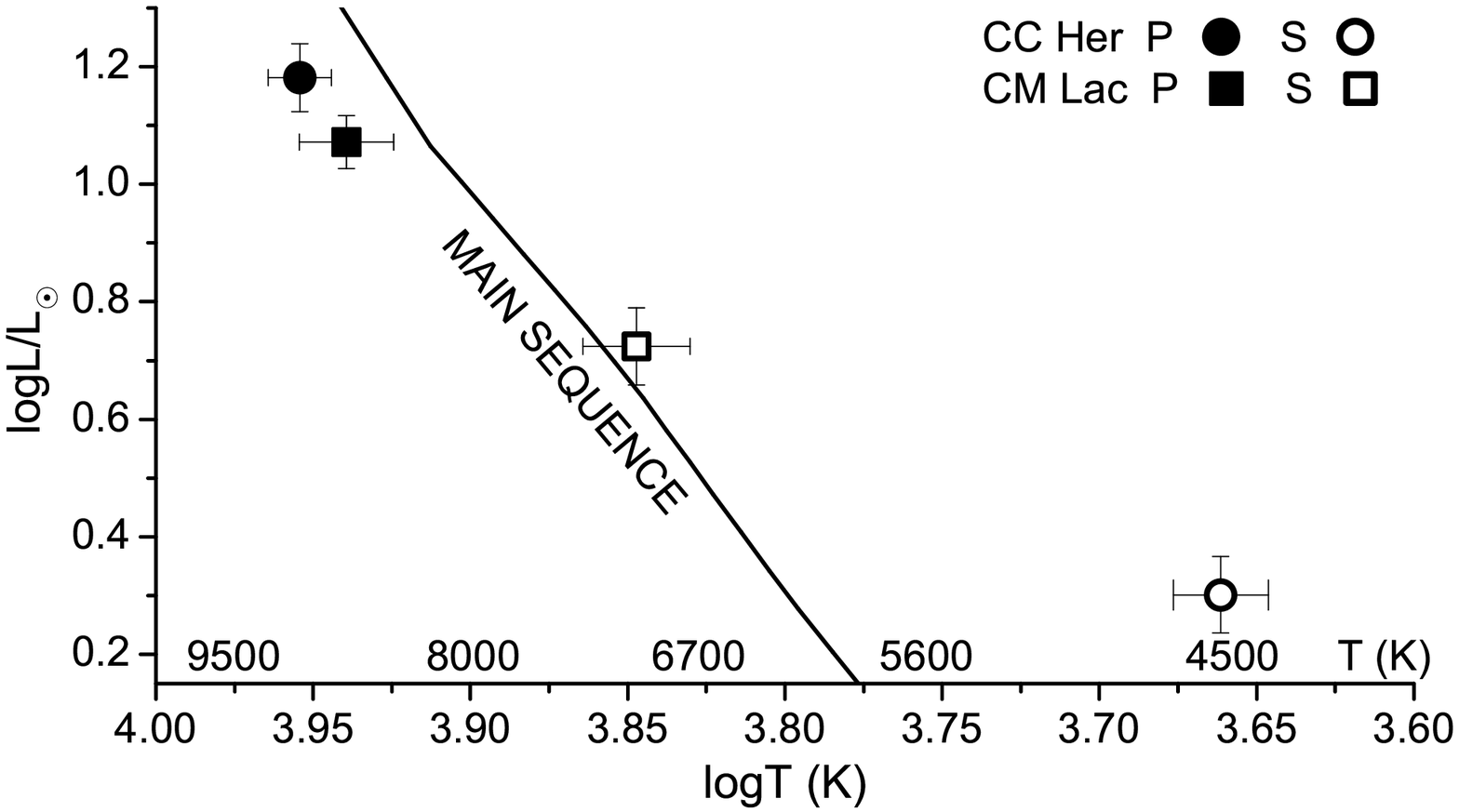}
\end{tabular}
\caption{The position of the components (P for primary and S for the secondary) of CC~Her and CM~Lac in $M-R$ (upper) and CM (lower) diagrams.}
\label{fig4}
\end{figure}

The geometric and photometric elements, derived from the simultaneous analysis of the light and RV curves, were used to compute the absolute elements of the components, which are listed in Table~\ref{tab4}. The temperatures of the components were assigned values resulted from the best fit model for each case (see Table~\ref{tab3} -- smallest $\Sigma res^2$). However, since the errors of the secondaries' temperatures are formal ones, we derived, according to the error propagation method, more realistic values. For that, the mean luminosity contribution of the secondary in the system's total luminosity (see Table~\ref{tab3}), the absolute radii of both components and an assumed, but very realistic, error of 300~K for $T_1$ (i.e. the temperature difference between two spectral subclasses close to the primary's spectral type) were used. Finally, these parameters were used to place the components on the Mass-Radius ($M-R$) and Colour-Magnitude (CM) diagrams (Fig.~\ref{fig4}) in order to examine their evolutionary status. The values for the Zero Age Main Sequence (ZAMS) and Terminal Age Main Sequence (TAMS) lines were taken from \citet{NM03}.

The secondary component of CC~Her was found to be at the subgiant stage of evolution, while the primary is still in the nucleus hydrogen-burning phase. This configuration is widely known as classical Algol type. On the other hand, both components of CM~Lac lie inside TAMS-ZAMS limits, revealing their MS nature. However, the primary component is closer to ZAMS line, while the secondary one is almost in the middle. These results are further discussed in the last section.

\begin{table}[t]
\centering
\caption{The absolute parameters of the systems' components (P: Primary, S: Secondary) with their respective errors corresponding to the last digit in parentheses.
\label{tab4}}
\begin{tabular}{lcc}
\tableline
Parameter	       &	\multicolumn{2}{c}{CC Her}	\\
\tableline
	               &	      P	    &	    S	    \\
\tableline
$M~(M_\odot$)	   &	1.91 (6)	&	0.56 (4)	\\
$R~(R_\odot$)	   &	1.61 (2)	&	2.27 (3)	\\
$T_{\rm eff}$ (K)  &9000 (300)$^a$	&	4586 (160)	\\
$\log g$ (cm/s$^2$)&	4.31 (2)	&	3.47 (3)	\\
$L~(L_\odot$)	   &	 15.2 (9)   &	2.0 (3)	    \\
$M_{\rm bol}$ (mag)&	1.8 (2)	    &	4.0 (2)	    \\
$a~(R_\odot$)	   &	1.85 (4)	&	6.35 (7)	\\
\tableline
                   &   \multicolumn{2}{c}{CM Lac}	\\
\tableline
$M~(M_\odot$)	   &	1.98 (6)	&	1.50 (10)	\\
$R~(R_\odot$)	   &	1.51 (3)	&	1.55 (3)	\\
$T_{\rm eff}$ (K)  &8700 (300)$^a$	&	7034 (270)	\\
$\log g$ (cm/s$^2$)&	4.37 (2)	&	4.23 (3)	\\
$L~(L_\odot$)	   &	11.8 (8)	&	5.3 (8)	    \\
$M_{\rm bol}$ (mag)&	2.1 (2)	    &	2.9 (2)	    \\
$a~(R_\odot$)	   &	3.77 (6)	&	4.97 (6)	\\
\tableline
\multicolumn{3}{l}{$^a$assumed}
\end{tabular}
\end{table}

\section{Search for pulsations}
\label{PULS}
CC~Her was listed by \citet{SOE06} as candidate to contain a pulsating component. In that study the only candidacy criterion was the spectral type of the components (A-F). Therefore, since CM~Lac is also an A-type binary, its LCs were tested for pulsations too. For this, the theoretical LCs in all filters were subtracted from the observations and a frequency search was performed in the interval 5 to 80~c/d \citep[typical for $\delta$ Sct stars,][]{BR00} in the out-of the primary eclipse data. The software \emph{PERIOD04} v.1.2 \citep{LB05}, that is based on classical Fourier analysis, was used. The results showed that neither star showed definite pulsational behaviour in the selected range of frequencies with a signal--to--noise ratio higher than 4, given the time resolution and the photometric errors provided for each system (see section 2).

\section{Discussion and conclusions}
\label{DIS}
New spectroscopic and multicolour photometric observations of CC~Her and CM~Lac were obtained and analysed with the most modern available techniques. The combination of the results derived from these methods made it possible to determine the absolute parameters of the components of the systems and allowed us to make a good estimate of their present evolutionary status. In addition, the four colour LCs provided the means for an accurate modelling and more cautious investigation for tertiary components by checking their possible light contribution to different filter data. The LCs of both systems were checked for pulsations, but the results were negative.

CC~Her is a classical Algol-type binary with its secondary component filling its Roche Lobe. Its primary component is a MS star, while the secondary is at the subgiant stage of evolution. The orbital period analysis of the system \citep{SOF06} did not show evidence of secular change of the period that can be connected with the mass transfer \citep[cf.][]{HI01}, although the system is semidetached. This discrepancy can be explained by assuming a very slow mass transfer rate that cannot be detected in the given date range of the O$-$C data.

In the same study a third body with minimal mass $\sim2.69~M_\odot$ was suggested as the most possible explanation for the cyclic period modulation. However, that mass value was based on an assumed total mass of 5.24~$M_\odot$ of the binary's components. Our results show that the components have masses of 1.91~$M_\odot$ and 0.56~$M_\odot$, for primary and secondary, respectively, hence the total mass $M_{1,2}$ is 2.47~$M_\odot$. Therefore, according to the results of \citet{SOF06}, considering the mass function value ($f(m_3)$=0.31~$M_\odot$) and using its formula \citep[cf.][]{TO10}:
\begin{equation}
f(M_3) = \frac{(M_3 \sin i_3)^3}{(M_1+M_2+ M_3)^2}
\end{equation}
we found that the minimal mass $M_{3,\rm min}$ (for $i_3=90^\circ$) of the potential third body is $1.77~M_\odot$. A close star with such mass is expected to be easily traced in both photometric and spectroscopic observations. In particular, assuming its MS nature, we can estimate its luminosity according to the Mass-Luminosity relation ($L \sim M^{3.5}$) for dwarf stars and compare it with the luminosity of the binary's components (see Table~\ref{tab4}) by using the following formula:
\begin{equation}
L_{3,\rm O-C} (\%)=100 \frac{M_{3,\rm min}^{3.5}}{L_1+L_2+M_{3,\rm min}^{3.5}}
\end{equation}
The above calculations resulted in a luminosity contribution of $\sim$30\%. On the other hand, the LC analysis revealed the existence of a third light contribution to the total luminosity, but its fraction was found to have a relatively small value ($\sim$1.7\%). Moreover, the RVs analysis identified only two peaks and no evidence of another third one. This large difference between the expected and observed luminosity results rejects the simplest scenario of a MS star orbiting the eclipsing pair. Very probably, one explanation for this could be the situation where the additional body is not a single star, but a multiple system that consists of low temperature components with a total mass of $1.77~M_\odot$. Alternatively, the absence of significant light contribution of the third body in the observed bands might allow us to think also that this body might be a compact object (e.g. neutron star or stellar black hole). Assuming a simultaneous birth of the components of the triple system and taking into account both the mass value of the third body and that the eclipsing pair has an evolved component, we conclude that the third body has had sufficient time to evolve into a compact object. Probably, future observations in another wavelengths (e.g. x-rays) might solve this mystery. However, for both previous scenarios concerning the nature of the third body, the expected luminosity contribution is much less than the calculated one for a single MS star and its absence from the spectra is reasonable.
Another explanation for this discrepancy could be the scenario of magnetic influences to the orbital period (quadrupole moment variation--$\Delta Q$) from the secondary component, widely known as the Applegate's mechanism \citep{AP92}. \citet{LR02} and \citet{RO00} proposed the relations for calculating the $\Delta Q$ of a binary's component. Therefore, by substituting to these relations the results of the O$-$C diagram analysis of \citep{SOF06}, and the radius of the secondary component and the system's major axis (see Table~\ref{tab4}) we found $\Delta Q\sim7.2\times10^{50}$~g~cm$^2$, which lies between the range $10^{50}<\Delta Q<10^{51}$ and could potentially implicate the cyclic period modulation of the system \citep{LR02}. Although we found that the secondary component is magnetically active (see Section~3.2), we cannot be certain that the Applegate's mechanism takes place in this system, since it also suggests brightness changes that cannot be checked from our data set due to the short time span of the observations. Concluding, astrometric observations for detecting motion around the barycenter of the possible multiple system and several years monitoring of the binary for brightness changes seem to be necessary for clarifying the multiplicity of CC~Her.

CM~Lac consists of two MS stars with almost same radii and different masses in a detached configuration. However, the less massive component appears slightly more evolved than the primary. This result comes in disagreement with the theory of stellar evolution considering the initial mass and assuming a simultaneous birth. Hence, this status feeds the evolutionary scenario of past mass transfer. Probably, the mass transfer was occurring in the past with a direction from the secondary to the primary component and, therefore, the secondary had been evolved faster than the primary, till the mass flow ended. However, now the components seem to be at relaxation phase and also in asynchronous orbit, something which is very common in detached systems. The present results for absolute parameters are in very good agreement with those of past studies \citep{PO68,LI73}. Contrary to that, our LCs and RVs analyses results do not show evidence for a tertiary component, as suggested by \citet{AA76} and \citet{HO06}.

Finally, CC~Her and CM~Lac can be considered as two very good `astrophysical labs' since they are both eclipsing double-line spectroscopic binaries and they provide a lot of useful information about the evolution of binary systems.

\section*{Acknowledgments}
This work has been financially supported by the Special Account for Research Grants No 70/4/11112 of the National \& Kapodistrian University of Athens, Hellas. Skinakas Observatory is a collaborative project of the University of Crete, and the Foundation for Research and Technology-Hellas. We thank Mr. Kougentakis and Dr. Semkov for the night assistance at Skinakas observatory, and the anonymous reviewer for the valuable comments that
improved the quality of the paper. In the present work, the SIMBAD database, operated at CDS, Strasbourg, France, and Astrophysics Data System Bibliographic Services (NASA) have been used.

\end{document}